\documentclass[graybox]{svmult}

\usepackage{helvet}         
\usepackage{courier}        
\usepackage{type1cm}        
\usepackage{makeidx}         
\usepackage{graphicx}        
\usepackage{multicol}        
\usepackage[bottom]{footmisc}
\usepackage{amsmath}
\usepackage{tabularx}

\begin{document}

\title*{Searching for discrete series representations at the late-time boundary of de Sitter}
\titlerunning{discrete series representations} 
\author{Gizem Şengör}
\institute{Gizem Şengör \at Physics Department, Boğaziçi University\\
34342 Bebek, İstanbul Turkey, \email{gizem.sengor@bogazici.edu.tr}
}
\maketitle

\abstract{The group $SO(d+1,1)$ makes an appearance both as the conformal group of Euclidean space in $d$ 
dimensions and as the isometry group of de Sitter spacetime in $d+1$ dimensions. While this common 
feature can be taken as a hint towards holography on de Sitter space, understanding the representation 
theory has importance for cosmological applications where de Sitter spacetime is relevant. Among the 
categories of $SO(d+1,1)$ unitary irreducible representations, discrete series is important in physical 
applications because they are expected to capture gauge fields. However, they are also the most difficult 
ones to recognize in field theoretical examples compared to representations from the other categories. 
Here we point towards some examples where we are able to recognize discrete series representations 
from fields on de Sitter and highlight some of the properties of these representations.}

\section{Introduction}
\label{sec:1 Intro}

The group $SO(d+1,1)$, lies at the intersection of multiple disciplines. Being the conformal group of Euclidean space in $d$ dimensions it is of interest to Euclidean Conformal Field Theory (CFT) \cite{Dobrev}. Being the isometry group of de Sitter spacetime in $d+1$ dimensions it is the group to address when discussing quantum field theory (QFT) on de Sitter \cite{QFTondS,unitarity,twopoint,discrete}. Having both a CFT and a curved spacetime side, makes $SO(d+1,1)$ also appear in Holography \cite{holography}. $SO(d+1,1)$ is a group about both Euclidean space and de Sitter spacetime. 

de Sitter spacetime itself is one of the three maximally symmetric vacuum solutions to the Einstein equations with a cosmological constant. This makes it a very intriguing question to compare how our methods developed for one of the other maximally symmetric vacuum spacetimes with a cosmological constant, be it in the context of QFT or CFT or Holography, work or need to be enlargened to address de Sitter spacetime. Physically de Sitter spacetime expresses accelerated expansion. As we are aware of two epochs of accelerated expansion in the cosmic history of our universe, the primordial inflationary epoch, and the current day dark energy epoch, de Sitter spacetime enters discussions in a very physical and observable setting; in Cosmology. From a broader perspective it is a group of interest to CFT, QFT, Holography and Cosmology. All of which are disciplines of interest both in Physics and Mathematics literature and which aim to address matter and gravity.

Wigner's identification of elementary particles on flat spacetime in terms of the unitary irreducible representations of Poincaré group \cite{Wigner}, the isometry group of Minkowski, has given us a solid framework to discuss quantum field theory and particle physics. Therefore to talk about the particles on de Sitter, which play a role both in our understanding of gravity and of observations, we turn on to the unitary irreducible representations of $SO(d+1,1)$. These fall under four categories: \emph{principal series, complementary series, exceptional series} and \emph{discrete series}. Here we are interested in unitary, irreducible symmetric traceless tensor representations, which are also referred to as type-I. 

From a physical perspective we tend to think of fields either in terms of matter fields or gauge fields which mediate interactions. Among the four categories of representations the \emph{discrete series representations} are expected to capture the gauge fields \cite{Guille}. As gravity is an interaction, learning about the properties of discrete series representations on de Sitter is part of the long winded goal of learning about gravity on de Sitter. At a more practical level, earlier on in \cite{unitarity} we were able to provide a list of principal and complementary series scalar operators at the late-time boundary of de Sitter, in a setting convenient for inflationary studies. Now we would like to enlarge this list by including in discrete series late-time operators. From a mathematical perspective, while each one of the four categories of representations have their individual properties, the discrete series representations have certain properties that make them even more unique then the rest. For instance discrete series representations, exist only when the rank of the group equals the rank of the maximally compact subgroup and for the group $SO(d+1,1)$ they can only be accommodated in terms of symmetric traceless tensors for two specific number of dimensions, $d+1=\{2,4\}$. There are no such restrictions on the dimensionality for the other categories \cite{Dobrev}. Here we focus on the discrete series representations for $d=1$ with the goal to recognize them in field theoretic examples. We will first discuss the unitarity properties and construct discrete series operators both in the bulk and on the late-time boundary. Motivated by the results of \cite{discrete}, we will focus on building highest and lowest weight states of a free massless scalar, at the late-time boundary of $dS_2$. This can further be considered in the context of BF theories for a more complete picture.

\section{Summary of subgroups of $SO(d+1,1)$ and Unitarity}
\label{sec:2 summary of reps}

The group $SO(d+1,1)$ is composed of the following subgroups: rotations $M=SO(d)$, dilatations $A=SO(1,1)$, special conformal transformations $N$, maximally compact subgroup $K=SO(d+1)$ and spatial translations $\tilde{N}$. What sets de Sitter in $d+1$ dimensions apart from Minkowski in $d+1$ dimensions is that this list does not involve time translations which exist for the Poincaré group, $ISO(d,1)$. Instead it has dilatations and special conformal transformations. The lack of time translations sets de Sitter apart from Anti de Sitter, whose isometry group in $d+1$ dimensions is $SO(d,2)$, as well.

In general, unitary irreducible representations are induced by the invariant subgroup. For the group $SO(d+1,1)$ there are two sets of invariant subgroups. One of them is made up of dilatations, special conformal transformations and rotations $(NAM)$. This invariant subgroup induces the principal and complementary series representations. The second invariant subgroup is the maximally compact subgroup $K$, and this is the one that induces discrete series representations.  Being induced by a compact group also hints the possibility to talk about highest and lowest weight states. 

The representation theory of the de Sitter group has been well studied in Mathematics literature, dating back to the works of Harisch-Chandra \cite{HC}. There are two main ways to construct representations, either from the action of finite group elements \cite{Dobrev}, or from the algebra \cite{Anous, Barut}. Recent reviews of both methods catered towards a CFT construction can be found in \cite{Sun review} and of the first method catered towards cosmological applications in \cite{unitarity}.  In global coordinates, as well as static patch coordinates, the metric of $dS_{d+1}$ can be analytically continued into the sphere metric $S^{d+2}$. Similarly, the group $SO(d+1,1)$ can be analytically continued to the group $SO(d+2)$. One can make use of this analytic continuation to capture the representations as well, where some of the generators are effected by the analytical continuation. This method has been employed to study scalars \cite{Higuchi analyt} a well as fermions \cite{Letsios} on de Sitter, and it works differently in $d+1=2$ dimensions \cite{Barut, Mukunda}, then it does in higher dimensions.  The unitary irreducible representations are labeled by the eigenvalues of the quadratic Casimir which involve \emph{spin} $s$ and \emph{scaling weight} $c$. The spin label is in general the eigenvalue of the rotation subgroup except for some of the exceptional series categories. We will denote these labels as $\chi=\{s,c\}$. For $d=1$, the rotation subgroup $M=SO(1)$ is trivial and the concept of spin is replaced with the concept of being odd or even \cite{Graev}. Among the two labels, the scaling weight plays a special role as it determines the categorization of the representations. The scaling weight is a part of the scaling dimension that carries information about the mass of the field. Under dilatations an operator with scaling dimensions $\Delta$ transforms as follows
\begin{equation}
\text{as}~\vec{x}\to\lambda \vec{x},~~\mathcal{O}(\lambda \vec{x})\to\lambda^{-\Delta}\mathcal{O}(\vec{x}),
\end{equation}
where $\vec{x}\in{R}^d$. For the group $SO(d+1,1)$ the scaling dimension has a very fixed format based on the number of spatial dimensions $d$ and the scaling weight $c$ as follows
\begin{equation}\label{intro c} \Delta=\frac{d}{2}+c.\end{equation}
For $SO(d+1,1)$ both real and purely imaginary $c$ give rise to unitary representations. However the range of $c$ determines the category of the representation. For comparison complex scaling dimensions of $SO(d,2)$ do not capture unitary representations, while complex scaling dimensions such that the scaling weight is the imaginary part, are unitary for $SO(d+1,1)$, these are the category of principal series, and correspond to heavy fields on de Sitter. This difference between dS and AdS is due to the lack and presence of time translations in each one. For the category of discrete series representations the scaling dimension is a positive integer. 

In building representations from finite group elements, these representations act on function spaces with certain properties and equipped with a well defined inner product. When building representations from the algebra, the focus is on normalizable states and the generators act on states. Unitarity means the representations acting on the functions of the function space, or the generators acting on states should preserve the well defined inner product on the function and on the state respectively. This boils down to figuring out what is the corresponding bra state for a given ket state. Here subtleties arise due to dilatations and this is why the range of the scaling weight works into the definition of unitarity. While standard Hermitian adjoint of the ket gives the bra in the case of principal series, the inner product needs to be modified to include an invertible, normalizable intertwining operator that carries on a similarity transformation such that $c\to -c$ for the rest of the categories. The intertwining operator being well defined depends on the range of the scaling weight and for this reason each representation category is equipped with its own set of intertwining operators. One can consult to \cite{Dobrev, Sun review} for the intertwining operators of various categories and to \cite{unitarity, twopoint, LT14} for the discussion focusing on scalar operators with examples that can be used in inflationary cosmology. In section \ref{subsec:bulk normalizaiton} we will review the discrete series inner product and hence the discrete series intertwining operator to the extend we need to make use of. A short review on the properties of the discrete series representations can also be found in \cite{Corfu}.

One can understand the necessity of different intertwining operators for different categories by considering the normalisation. A convenient choice for the normalization of complementary series intertwining operator is
\begin{equation}\label{compGnorm}
    n^+(\chi)=\left(\frac{d}{2}+s+c-1\right)\frac{\Gamma\left(\frac{d}{2}+c-1\right)}{\Gamma\left(-c\right)}.
\end{equation}
However note that the Gamma function $\Gamma(z)$, has poles for $z=0,-1,-2,\dots$.
In general for scalar fields $c=\pm\sqrt{\frac{d^2}{4}-m^2l^2}$ and for a massless scalar we have $c=\pm\frac{d}{2}$ which is captured by exceptional series. When $d=\{1,3\}$, the normalization becomes ill defined for $c=-\frac{d}{2}$, which is  where we expect to find discrete series representations for $d=\{1,3\}$. The exceptional series are in general reducible into discrete series \cite{Dobrev} and when $d=\{1,3\}$ they overlap with discrete series \cite{Guille}.

The function spaces on which the representations act are labeled by spin and scaling weight, just like the representations. In talking about normalization one needs to pay attention to which well defined intertwining operator maps which two functions spaces. For the case of scalars on $d=1$ parity is trivial and it is enough to label the function space by scaling dimension. The function spaces $\mathcal{C}_{1-\Delta}$ and $\mathcal{C}_{\Delta}$ with $\Delta=1,2,3,\dots$ are the ones that capture discrete series representations that we are after. 

In the next subsection we will obtain both bulk and late-time operators in a specific example, identify which function space they belong to and normalize them. In section \ref{sec:highestlowest} we will build highest and lowest weight states from some of the late-time operators.

\section{Identifying scalar discrete series operators at the late-time boundary of $dS_2$}

 We will work in conformal global coordinates, where the $dS_2$ metric takes the following form
\begin{equation}\label{metric} \frac{ds^2}{l^2}=\frac{-dT^2+d\theta^2}{\sin^2T},~~T\in(-\pi,0)
\end{equation}
and $l$ is the de Sitter radius. The generators can be realized as differential operators in terms of the Killing vectors. We will write them as in \cite{discrete}
\begin{subequations}
\label{eq:diff ops}
\begin{align}
\label{eqn:L0}L_0&=-i\partial_\theta,\\
\label{eqn:L+1}L_{+1}&=-e^{-i\theta}\left(i\cos T\partial_\theta+\sin T\partial_T\right),\\
\label{eqn:L-1}L_{-1}&=-e^{+i\theta}\left(i\cos T\partial_\theta-\sin T\partial_T\right).
\end{align}
\end{subequations}
The generator $L_0$ is the generator of the maximally compact subgroup, which in this case is $K=SO(2)$. These generators obey the $so(2,1)$ algebra, whose double cover is the $sl(2,R)$ algebra, which can be written as
\begin{equation}\label{algebra}\left[L_m,L_n\right]=(m-n)L_{n+m}.\end{equation}

We introduce our free massless scalar field as is customary for a real scalar field in quantum field theory, by expanding it in terms of mode functions and annihilation and creation operators such that the expression is real. In the basis of the maximally compact subgroup generator we have
\begin{equation}\label{phiexp}
    \phi(T,\theta)=\sum_{n\in Z/\{0\}}\frac{e^{in\theta}}{2\pi}\left[\phi_n(T)a_n+\phi^*_{-n}(T)a^\dagger_{-n}\right].
\end{equation}
One can think of label $n$ as a discrete momentum label, and the expansion \eqref{phiexp} is similar to going to momentum space often employed in inflationary cosmological calculations. The mode functions satisfy the equation of motion
\begin{equation} \partial^2_T\phi_n+n^2\phi_n=0.\end{equation}
Here we remove the zero mode ($n=0$) by hand, which we will comment on more later on. However one can make use of the gauge invariance of the massless scalar or consider it in the context of a BF theory, which is a gauge theory with a scalar and a vector field, to put the removal of the zero mode in a more rigorous context \cite{discrete}. We consider the field to be quantized and demand $a_n$ and $a^\dagger_n$ be the annihilation and creation operators satisfying
\begin{equation}
  \label{aadag} \left[a_n,a^\dagger_m\right]=\delta_{n,m},~~
a_n|0\rangle=0,~~a^\dagger_n|0\rangle=|\Delta,n\rangle,~~\langle 0|a_n=\langle\Delta,n| 
\end{equation}
with $\delta_{n,m}$ being the Kronecker delta, the annihilation operator $a_n$ annihilates the de Sitter invariant vacuum state $|0\rangle$ and label $n$ stands for the $L_0$ eigenvalue, which we will comment more on in the next section. The general solution for the mode functions leads to
\begin{equation}\label{eqn:bulkphi}
  \phi(T,\theta)=\sum_{n\in Z/\{0\}}\frac{e^{in\theta}}{2\pi}\left[\left(A \cos nT+ B\sin nT\right)a_n+\left(A^* \cos nT-B^*\sin nT\right)a^\dagger_{-n}\right].   
\end{equation}
Our goal is to decompose the field in terms of operators which we can recognize as the unitary irreducible representations of $SO(2,1)$. We will follow two complementary routes, first we will focus on operators in the bulk\footnote{We thank the anonymous referee for pointing out to us this way of organizing the field. The discussion in the following subsection \ref{subsec:bulk decomposition} has evolved mainly by the referee's input.}, next we will focus on the late-time limit and make use of the holographic dictionary.

\subsection{The bulk decomposition}
\label{subsec:bulk decomposition}
 The abstract generators, which we will denote as $\hat{L}_m$ following \cite{discrete}, and the Casimir $\hat{\mathcal{C}}$ of the algebra \eqref{algebra} acts on states $|\Delta,n\rangle$ defined in \eqref{aadag} as follows
  \begin{equation}\label{genonst}
\hat{\mathcal{C}}|\Delta,n\rangle=\Delta(\Delta-1)|\Delta,n\rangle,~\hat{L}_0|\Delta,n\rangle=-n|\Delta,n\rangle,~\hat{L}_{\pm 1}|\Delta,n\rangle=-(n\pm\Delta)|\Delta,n\pm1\rangle.\end{equation}
In the bulk decomposition, we want to focus on the mode functions and make sure that they explicitly realize unitary irreducible representations of $SO(2,1)$. For this purpose, it is convenient to start first by labeling them with $\Delta,n$ just like the states. Thus we introduce $\phi^{(\Delta)}_n(T,\theta)$ such that they obey\footnote{We thank the anonymous referee for pointing out to us to organize the modefunctions this way.}
\begin{equation}\label{eq:modefuncintroduce}
\hat{L}_m\phi^{(\Delta)}_n(T,\theta)=-\left(n+m\Delta\right)\phi^{(\Delta)}_{n+m}(T,\theta)~~\text{for}~~\Delta=0,1.\end{equation}
With this purpose, choosing $A=-B=\frac{1}{\sqrt{2}}$ and replacing\footnote{This replacement of $n\to-n$ is necessary to have the modefunctions satisfy the algebra as stated in \eqref{eq:modefuncintroduce}.} $n$ with $-n$  in the sum, we can rewrite \eqref{eqn:bulkphi} as
\begin{align}
  \label{eqn:bulkphi_reorganized} \phi(T,\theta)=\sum_{n\in Z/\{0\}}\left[\phi^{(0)}_n(T,\theta)~^B\alpha_n+\phi^{(1)}_n(T,\theta)~^B\beta_{-n}\right]
\end{align}
with the modefunctions being
\begin{equation}
    \label{eqn:UIRmodes}
        \phi^{(0)}_n(T,\theta)=\frac{e^{-in\theta}}{2\pi}\cos{nT},~~
        \phi^{(1)}_n(T,\theta)=\frac{e^{-in\theta}}{2\pi}\frac{\sin{nT}}{n},
\end{equation}
and the bulk operators defined as
\begin{equation}
    \label{n_spaceops_intro}  
~^B\alpha_n=\frac{1}{\sqrt{2}}\left[a_{-n}+a^\dagger_n\right],~~
 ~^B\beta_n=\frac{n}{\sqrt{2}}\left[-a_{n}+a^\dagger_{-n}\right].
\end{equation}
Equation \eqref{aadag} implies that the bulk operators $~^B\alpha_n$ and $~^B\beta_n$ obey the following nontrivial commutation relation
\begin{equation}\label{eqn:alpha_nbeta_mcomm}
    \left[~^B\alpha_n,~^B\beta_m\right]=n\delta_{n,m}.
\end{equation}

As a check, action of the differential representations \eqref{eq:diff ops} on \eqref{eqn:UIRmodes} does satisfy \eqref{eq:modefuncintroduce}. Comparing the action of the differential operators \eqref{eqn:L+1} and $\eqref{eqn:L-1}$ on $\phi^{(0)}_n(T,\theta)$ and $\phi^{(1)}_n(T,\theta)$ with \eqref{eq:modefuncintroduce}, one can confirm that $\phi^{(0)}_n(T,\theta)$ has dimension $\Delta=0$, implying that it corresponds to a state in the function space $\mathcal{C}_0$ and $\phi^{(1)}_n(T,\theta)$ has dimension $\Delta=1$, making it a state in $\mathcal{C}_1$. In other words the mode functions $\phi_n$ in \eqref{phiexp} that obey the equations of motion, are also related to $\phi^{(\Delta)}_n$, the eigenfunctions of $L_0$.

At this point, for $\phi^{(1)}_n$ it is necessary to exclude $n=0$ where as it can be kept for $\phi^{(0)}_n$. However we will continue to remove the $n=0$ mode as we have done so by hand in the sum in \eqref{eqn:bulkphi_reorganized}.

\subsection{Normalization of $~^B\alpha_n$ and $~^B\beta_n$}
\label{subsec:bulk normalizaiton}
Next we wish to identify which function space the operators $~^B\alpha_n$ and $~^B\beta_n$ belong to in order to normalize them. As in the previous section, we can make use the action of $\hat{L}_{\pm1}$ to read off the dimension. So far we have worked out how $L_{\pm1}$ acts on $\phi(T,\theta)$ via differential operators \eqref{eq:diff ops} and how it acts on the mode functions $\phi^{(\Delta)}_n(T,\theta)$ via \eqref{eq:modefuncintroduce}. Using the action of the generators on the modefunctions, we can work out the action of the generators on the operators $~^B\alpha_n$, $~^B\beta_n$. Thus we introduce $\mathcal{L}^{(\Delta)}_m$ through 
\begin{equation}
    -L_m\phi(T,\theta)=\sum_n\left\{\phi^{(0)}_n(T,\theta)\left[\mathcal{L}^{(0)}_m,~^B\alpha_n\right]+\phi^{(1)}_n(T,\theta)\left[\mathcal{L}^{(1)}_m,~^B\beta_{-m}\right]\right\}
\end{equation}
where
\begin{equation}\label{eqn:mathcalLcomm}
    \left[\mathcal{L}^{(0)}_m,~^B\alpha_n\right]=(n-m)~^B\alpha_{n-m}~~\text{and}~~\left[\mathcal{L}^{(1)}_m,~^B\beta_{-n}\right]=n~^B\beta_{-n+m}.
\end{equation}
The commutators in \eqref{eqn:mathcalLcomm} are achieved by making use of \eqref{eq:modefuncintroduce} and rearranging the indices appropriately. 
Focusing on $m=1$ in \eqref{eqn:mathcalLcomm} we have
\begin{equation}\label{eqn:dim alpha_n beta_n}
    \left[\mathcal{L}^{(0)}_1,~^B\alpha_{-n}\right]=-(n+1)~^B\alpha_{-(n+1)}~~\text{and}~~\left[\mathcal{L}^{(1)}_1,~^B\beta_{-n}\right]=n~^B\beta_{-n+1}.
\end{equation}
Comparing the above with the action of $\hat{L}_m$ in \eqref{genonst}, we conclude that $~^B\alpha_{-n}=~^B\alpha^\dagger_n$ is a $\Delta=1$ operator with respect to $\hat{L}_m$. In other words $~^B\alpha_n$ is a $\Delta=1$ operator in the representation of $\hat{L}_m^\dagger=-\hat{L}_{-m}$. Similarly $~^B\beta_n$ is a $\Delta=0$ operator among the representations of $\hat{L}_m$.

Lastly, we can write the generators $\mathcal{L}^{(\Delta)}_m$ that we have introduced in \eqref{eqn:mathcalLcomm} in terms of the operators $~^B\alpha_n$, $~^B\beta_n$. Taking into account the commutation relation \eqref{eqn:alpha_nbeta_mcomm}, 
\begin{equation}
    \label{eqn:mathcalL} \mathcal{L}^{(\Delta)}_m=-\sum_k\frac{k+m\Delta}{k+m}~^B\beta_{k+m}~^B\alpha_k
\end{equation}
realizes \eqref{eqn:mathcalLcomm}.

At this point, among the representations of $\hat{L}_m$ we have identified
\begin{subequations}\label{eqn:dim alpha beta}
 \begin{align} ~^B\alpha_{-n}&=\frac{1}{\sqrt{2}}\left[a_n+a^\dagger_{-n}\right]\in\mathcal{C}_1,\\
  ~^B\beta_n&=\frac{n}{\sqrt{2}}\left[-a_n+a^\dagger_{-n}\right]\in\mathcal{C}_0\end{align}
\end{subequations}
From these operators we define the normalized bulk operators, $~^B\alpha^N_{-n}=\mathcal{N}_\alpha~^B\alpha_{-n}$ and $~^B\beta^N_n=\mathcal{N}_\beta~^B\beta_n$ whose normalization we will now work out using the discrete series inner product. 

For dimension $\Delta$, there is the following position space intertwining operator \cite{Sun review, Dobrev, Graev}
\begin{equation}
    \label{eqn:intertwiner_pos} G^{(\Delta)}(x)=\frac{i}{2}\frac{1}{\Gamma(2\Delta)}\partial_x^{2\Delta-1}:\mathcal{C}_{1-\Delta}\to\mathcal{C}_\Delta.
\end{equation}
In $n$-space, this will be
\begin{equation}
    \label{eqn:intertwiner_n} G^{(\Delta)}(n)=\frac{i}{2\Gamma(2\Delta)}(-in)^{2\Delta-1}:\mathcal{C}_{1-\Delta}\to\mathcal{C}_\Delta.\end{equation}
For $\Delta=1$, regarding \eqref{eqn:dim alpha beta} the intertwining operator will act on $\alpha_n$ and $\beta_n$ as follows
\begin{equation}
    \label{eqn:int on alpha beta} ~^B\alpha_{-n}=G^{(1)}(n)~^B\tilde{\alpha}_{-n},~~~^B\tilde{\beta}_n=G^{(1)}(n)~^B\beta_n
\end{equation}
 with
 \begin{equation}\label{eqn:Gn}
     G^{(1)}(n)=\frac{n}{2}:\mathcal{C}_0\to\mathcal{C}_1.
 \end{equation}
 In normalizing the bulk operators, we first we identify the intertwined operators $\tilde{\alpha}_{-n}$ and $\tilde{\beta}_{n}$ using \eqref{eqn:Gn} in \eqref{eqn:int on alpha beta}. Then defining states from our operators as
\begin{equation}\label{building_st}|\mathcal{O}\rangle=\mathcal{O}_n|0\rangle,\end{equation}
  and the inner product as follows
  \begin{equation}
\left(\mathcal{O},\tilde{\mathcal{O}}\right)=\frac{1}{\Omega}\sum \langle\mathcal{O}|\tilde{\mathcal{O}}\rangle,~~\Omega\equiv\sum\langle\Delta,-n|\Delta,-n\rangle.
  \end{equation}
we arrive at the following normalized bulk discrete series operators
\begin{equation}
        \label{eqn:bulkNA NB}~^B\alpha_{-n}^N=\frac{\sqrt{|n|}}{\sqrt{2}}\left[a_n+a^\dagger_{-n}\right],~~
        ~^B\beta^N_n=(-i)\frac{\sqrt{2}}{\sqrt{|n|}}\left[-a_n+a^\dagger_{-n}\right].
\end{equation}
The appearance of factors of $n$ in the normalization carry on the information that $n=0$ mode is problematic, especially for $~^B\beta_n^N$, and is one justification of our removal of it. Lastly, notice that
\begin{equation}
    \left(~^B\alpha^N_{-n}\right)^\dagger=~^B\alpha^N_n,~~\left(~^B\beta\right)^\dagger=~^B\beta^N_{-n}.
\end{equation}

\subsection{Late-time decomposition}
\label{subsec:late-time decomposition}

From the point of view of holography we expect to have operators appear at the boundary of spacetime of interest. Moreover, the holographic dictionary  determines the scaling dimensions of the operators from the field behavior at the boundary. For de Sitter there are two spacelike boundaries: one at early time and one at late-time. Here we focuse on the boundary at late-time. Employing the holographic dictionary via the late-time limit determines the scaling dimensions from the time dependence of the field behaviour at late-times. Considering the holographic dictionary in general dimensions in the case of scalars, we expect to find operators in the late-time limit with certain scaling dimensions appearing in the following format
\begin{equation}
    \label{eqn:holodic}\lim_{T\to0}\phi(T,\theta)= T^{\bar{\Delta}}\mathcal{O}_{\bar\Delta}(\theta)+T^{\Delta}\mathcal{O}_{\Delta}(\theta),
\end{equation}
where $\Delta+\bar\Delta=d$. 


Our goal is to obtain the operators $\mathcal{O}_\Delta$ and $\mathcal{O}_{\bar\Delta}$, in a way we can recognize them as unitary irreducible representations of the de Sitter group $SO(2,1)$. In doing so, our key tool will be to identify which function space they belong to and be able to normalize them as is expected for the representation category.

Keeping the convention that $A=-B=\frac{1}{\sqrt{2}}$, the field \eqref{eqn:bulkphi} is reorganized in the late-time limit with respect to the holographic dictionary \eqref{eqn:holodic} as 
\begin{equation}\label{eqn:latetimelim}\lim_{T\to0}\phi(T,\theta)=\sum_{n\in Z/\{0\}}\frac{e^{in\theta}}{2\pi}\left[~^{lt}\alpha_n+T ~^{lt}\beta_n\right]\end{equation}
with
\begin{equation}
    ~^{lt}\alpha_n=\frac{1}{\sqrt{2}}\left(a_n+a^\dagger_{-n}\right),~~~^{lt}\beta_n=\frac{n}{\sqrt{2}}\left(-a_n+a^\dagger_{-n}\right).
\end{equation}
Reading off the dimensions by comparison from the holographic dictionary we expect to find the following two position space operators
\begin{subequations}\label{thetasp_LTops}
    \begin{align}
        ~^{lt}\alpha(\theta)&=\sum_{n\in Z/\{0\}}\frac{e^{in\theta}}{\sqrt{2\pi}}~^{lt}\alpha_n,~~\Delta_\alpha=0,\\
        ~^{lt}\beta(\theta)&=\sum_{n\in Z/\{0\}}\frac{e^{in\theta}}{\sqrt{2\pi}}~^{lt}\beta_{n},~~\Delta_\beta=1.
    \end{align}
\end{subequations}
This let's us to identify the operators $~^{lt}\alpha_n$ and $~^{lt}\beta_n$ to belong to the following function spaces
\begin{equation}
    ~^{lt}\alpha_n\in\mathcal{C}_0,~~~^{lt}\beta_n\in\mathcal{C}_1.
\end{equation}
Then the intertwining operator and the normalization procedure we discussed in detail in section \ref{subsec:bulk normalizaiton} lead to the following normalized late-time operators
 \begin{equation}\label{Normalized latetime ops}
      ~^{lt}\alpha^N_n=\frac{\sqrt{2}}{\sqrt{|n|}}\left[a_n+a^\dagger_{-n}\right],~~^{lt}\beta^N_n=-i\frac{\sqrt{|n|}}{\sqrt{2}}\left[a_n-a^\dagger_{-n}\right]
  \end{equation}
  These discrete series operators also appear in the context of Abelian BF theory on $dS_2$ \cite{discrete} (here our convention is such that the operator with the lower scaling dimensions is $\alpha$, opposite to the convention of \cite{discrete} and our finding in section \ref{subsec:bulk decomposition}). Just like in the case of bulk operators factors of $n$ and where they appear in the normalization justify our removal of the zero mode. We would like to note two properties of our normalized discrete series late-time operators. Their Hermitian adjoints satisfy
  
  \begin{equation}\label{Hermadj} ~^{lt}{\alpha^N_n}^\dagger=~^{lt}\alpha^N_{-n},~~~^{lt}{\beta^N_n}^\dagger=~^{lt}\beta^N_{-n}.
  \end{equation}
  If we go back to position space
  \begin{equation}\mathcal{O}\equiv\sum \frac{e^{in\theta}}{\sqrt{2\pi}}\mathcal{O}_n,\end{equation}
  our normalized operators satisfy the following nontrivial commutation
  \begin{equation}\left[~^{lt}\alpha^N(\theta),~^{lt}\beta^N(\theta')\right]=2i\delta(\theta-\theta')
  \end{equation}
  similar to that of a quantized canonical field and its conjugate momentum up to a factor of 2.

  \section{Massless scalar highest and lowest weight states}
  \label{sec:highestlowest}
In the basis of maximally compact subgroup generator $L_0$, using our normalized late-time operators we can build states, as we did in \eqref{building_st}, which we also know are labeled by the Casimir eigenvalue $\Delta$ and the maximally compact subgroup eigenvalue $n$
\begin{equation}\label{state} |\mathcal{O}^N\rangle\equiv \mathcal{O}^N_n|0\rangle\equiv|\Delta,n\rangle.\end{equation}
Remembering the action of the abstract generators \eqref{genonst}, by definition the highest and lowest weight states get annihilated by the raising and lowering operators respectively
\begin{subequations}
    \begin{align}
        \text{Highest weight:}&~\hat{L}_{+1}|\Delta,n\rangle=0,\\
        \text{Lowest weight:}&~\hat{L}_{-1}|\Delta,n\rangle=0.
    \end{align}
\end{subequations}
Focusing on our normalized late-time operator $~^{lt}\beta^N_n$ we see that
\begin{equation}
    |1,-1\rangle\equiv ~^{lt}\beta^N_1|0\rangle
\end{equation}
is the highest weight state and
\begin{equation}
    |1,-n\rangle\equiv ~^{lt}\beta^N_n|0\rangle,~\text{for}~~n=1,2,\dots
\end{equation}
gives a tower of highest weight states. Similarly
\begin{equation}
    |1,1\rangle\equiv~^{lt}\beta^N_{-1}|0\rangle
\end{equation}
is the lowest weight state and gives rise to the following tower of lowest weight states
\begin{equation}
    |1,n\rangle\equiv~^{lt}\beta^N_{-n}|0\rangle,~\text{for}~~n=1,2,\dots
\end{equation}
Notice that since we removed the zero-mode the tower of highest weight states is disconnected from the tower of lowest weight states.

\section{Concluding remarks}
A characteristic feature of discrete series representations is that the operators obtained by intertwining maps are not unitarily equivalent. Physically this gives rise to a tower of highest weight states that are distinct from the tower of lowest weight states. We have seen this feature arise explicitly in our example in section \ref{sec:highestlowest}, and related it to the removal of the zero mode. In section \ref{sec:highestlowest} we made use of the late-time operators constructed in section \ref{subsec:late-time decomposition}. In addition to these we also discussed bulk operators in sections \ref{subsec:bulk decomposition} and \ref{subsec:bulk normalizaiton}. 

The generators $\hat{L}_n$ can also be expressed in terms of the annihilation and creation operators \cite{discrete}
\begin{equation}
    \hat{L}_n=-\sum_{k=-\infty}^\infty \left(k+n\Delta\right)a^\dagger_{k+n}a_k.
\end{equation}
We can invert \eqref{Normalized latetime ops} to rewrite the annihilation and creation operators in terms of our normalized discrete series late-time operators
\begin{equation}
        a_n=\frac{1}{2}\left[\frac{\sqrt{|n|}}{\sqrt{2}}~^{lt}\alpha_n^N+i\frac{\sqrt{2}}{\sqrt{|n|}}~^{lt}\beta^N_n\right],~~
        a^\dagger_n=\frac{1}{2}\left[\frac{\sqrt{|n|}}{\sqrt{2}}~^{lt}\alpha_{-n}^N-i\frac{\sqrt{2}}{\sqrt{|n|}}~^{lt}\beta^N_{-n}\right].
\end{equation}
Finally the generators can also be expressed in terms of the normalized discrete series late-time operators
\begin{eqnarray}
\nonumber    \hat{L}_n=-\frac{1}{4}\sum_{k=-\infty}^\infty &\left(k+n\Delta\right)\Bigg[\frac{\sqrt{|k||k+n|}}{2}~^{lt}\alpha^N_{-k-n}~^{lt}\alpha^N_k+\frac{2}{\sqrt{|k||k+n|}}~^{lt}\beta^N_{-k-n}~^{lt}\beta^N_k\\
&+i\sqrt{\frac{|k+n|}{|k|}}~^{lt}\alpha^N_{-k-n}~^{lt}\beta^N_k-i\sqrt{\frac{|k|}{|k+n|}}~^{lt}\beta^N_{-k-n}~^{lt}\alpha^N_k\Bigg].
\end{eqnarray}
By inverting equations \eqref{eqn:bulkNA NB} to rewrite annihilation and creation operators in terms of the normalized bulk operators we can also related the bulk and late-time operators. The relation comes out to be
\begin{equation}
    \label{bulk-latetime con}~^{lt}\alpha_{n}^N=\frac{2}{|n|}~^B\alpha^N_{-n},~~~^{lt}\beta_{n}^N=-\frac{|n|}{2}~^B\beta^N_n.
\end{equation}
We leave it for future to further discuss the relation between the bulk and the boundary operators.

 The treatment of the zero mode was a key point in our discussion. Removing the zero mode corresponds to removing one entire de Sitter representation and therefore does not spoil de Sitter invariance \cite{discrete}. Note that we have not argued whether the zero mode is a unitary or nonunitary representation. Earlier on in literature, the zero mode has been identified as a gauge mode \cite{Higuchi zeromodes}. This identification has been made more rigorous recently in a specific example of a spin-1 field \cite{Guille}. As was already mentioned, the removal of the zero mode can be made more rigourous in the context of interacting quantum field theories, where in the end the discrete series states do not end up in the physical Hilbert space once gauge invariance is taken into account, moreoever the removal of the zero mode has implications for locality \cite{discrete}. We leave it for future work to explore more the gauge theory side of the zero mode and its treatment for other spin and scaling dimensions. 

\begin{acknowledgement}
The author would like to thank Dionysios Anninos, Tarek Anous, Higuchi Atsushi, Vasileios Letsios, Guillermo A. Silva, Benjamin Pethybridge, Math$\acute{ı}$as Sempé, the participants of the LT15 workshop for helpful discussions and the anonymous referee for their helpful comments overall and input that led to the discussion on bulk operators. This work is funded by TÜBİTAK(The Scientific and Technological Research Council of Turkey) 2232 - B International Fellowship for Early Stage Researchers programme with project number 121C138. The author also thanks the editors for allowing extra time during the revision phase due to maternity leave and her newborn daughter and husband for creating the time to incorporate the revisions.
\end{acknowledgement}
\end{document}